\documentclass[12pt]{article}
\usepackage{graphicx}
\usepackage{epsfig}
\usepackage{epstopdf}
\DeclareGraphicsExtensions{.pdf,.eps,.png,.jpg,.mps}

\usepackage{cite}

\def\lsim{\mathrel{\rlap{\lower3pt\hbox{\hskip0pt$\sim$}}
     \raise1pt\hbox{$<$}}}         
\def\gsim{\mathrel{\rlap{\lower4pt\hbox{\hskip1pt$\sim$}}
     \raise1pt\hbox{$>$}}}         

\usepackage{amsmath}
\usepackage{amsfonts}

\begin{document}
\begin{titlepage}

\centerline{\Large \bf Decoding Stock Market with Quant Alphas}
\medskip

\centerline{Zura Kakushadze$^\S$$^\dag$\footnote{\, Zura Kakushadze, Ph.D., is the President of Quantigic$^\circledR$ Solutions LLC,
and a Full Professor at Free University of Tbilisi. Email: zura@quantigic.com} and Willie Yu$^\sharp$\footnote{\, Willie Yu, Ph.D., is a Research Fellow at Duke-NUS Medical School. Email: willie.yu@duke-nus.edu.sg}}
\bigskip

\centerline{\em $^\S$ Quantigic$^\circledR$ Solutions LLC}
\centerline{\em 1127 High Ridge Road \#135, Stamford, CT 06905\,\,\footnote{\, DISCLAIMER: This address is used by the corresponding author for no
purpose other than to indicate his professional affiliation as is customary in
publications. In particular, the contents of this paper
are not intended as an investment, legal, tax or any other such advice,
and in no way represent views of Quantigic$^\circledR$ Solutions LLC,
the website \underline{www.quantigic.com} or any of their other affiliates.
}}
\centerline{\em $^\dag$ Free University of Tbilisi, Business School \& School of Physics}
\centerline{\em 240, David Agmashenebeli Alley, Tbilisi, 0159, Georgia}
\centerline{\em $^\sharp$ Centre for Computational Biology, Duke-NUS Medical School}
\centerline{\em 8 College Road, Singapore 169857}
\medskip
\centerline{(April 25, 2017)}

\bigskip
\medskip

\begin{abstract}
{}We give an explicit algorithm and source code for extracting expected returns for stocks from expected returns for alphas. Our algorithm altogether bypasses combining alphas with weights into ``alpha combos". Simply put, we have developed a new method for trading alphas which does not involve combining them. This yields substantial cost savings as alpha combos cost hedge funds around 3\% of the P\&L, while alphas themselves cost around 10\%. Also, the extra layer of alpha combos, which our new method avoids, adds noise and suboptimality. We also arrive at our algorithm independently by explicitly constructing alpha risk models based on position data.\footnote{\, This is the last paper in the trilogy, which contains ``Factor Models for Alpha Streams" \cite{AlphaFM} and ``How to Combine a Billion Alphas" \cite{Billion}.} Forecasting stock returns with quant alphas has implications for the investment industry.
\end{abstract}
\medskip
\end{titlepage}

\newpage
\section{Introduction and Summary}

{}Not long ago quant trading workshops featured ``man v. machine" debates. Well, it is a foregone conclusion. Quantitative alpha\footnote{\, Here ``alpha" -- following the common trader lingo -- generally means any reasonable ``expected return" that one may wish to trade on and is not necessarily the same as the ``academic" alpha.  In practice, often the detailed information about how alphas are constructed may not even be available, e.g., the only data available could be the position data, so ``alpha" then is a set of instructions to achieve certain stock (or some other instrument) holdings by some times $t_1,t_2,\dots$} mining is now done by machines. Human's role in this process has essentially shifted to coding up various machine learning, data mining, clustering and other similar algorithms. Hardware is cheap, so mining millions of alphas is no longer a dream but the reality. Unsurprisingly, these exponentially proliferating alphas are ever fainter and more ephemeral.

{}A typical such alpha cannot even be traded on its own -- its signal is too weak to make any money after trading costs.\footnote{\, This includes transaction costs (exchange, broker-dealer, SEC, etc., fees) as well as slippage.} So, quant traders follow an ancient ``there is strength in numbers" wisdom and combine a large number of these faint alphas into a single ``mega-alpha" with some nontrivial weights. The game then becomes how to pick these weights optimally. This is essentially an {\em alpha portfolio} optimization problem. And it is a nontrivial one. Ostensibly, it is similar to a stock portfolio optimization problem \cite{Markowitz}, \cite{Sharpe94}. However, there is an important detail that makes all the difference: the number of alphas can be huge, in hundreds of thousands, millions or even billions. The available history (lookback), however, naturally is much shorter,\footnote{\, E.g., with daily observations, there are $T \sim 250$ datapoints in a (generous) 1 year lookback.} precisely due to the ephemerality of these alphas.\footnote{\, Additional considerations such as optimizing the turnover of the ``mega-alpha", trading which (as opposed to the individual alphas) offers an automatic benefit of internal crossing of trades (i.e., trading cost reduction), and scalability (i.e., how much capital this ``mega-alpha" can absorb before its market impact results in diminishing returns) add further complexity to the problem.}

{}There are various ways of approaching this problem of combining a large number of alphas. If the only available information is the time series of alpha returns, then the playing field is limited. As discussed in \cite{Billion}, modeling the alpha portfolio risk via a statistical risk model\footnote{\, For a recent discussion of statistical risk models, see \cite{StatRM}.} based solely on this time series or its extension via adding a few ``style" risk factors\footnote{\, E.g., turnover, etc. See \cite{AlphaFM} and \cite{Billion} for details.} at the end of the day invariably leads to a simple answer that the alpha weights are\footnote{\, Up to corrections suppressed by powers of $1/N$, where $N$ is the number of alphas.} proportional to residuals of a (weighted) regression. Simply put, it is the size of the available data (in this case, the lookback) that determines how much of the alpha risk space we can cover.

{}So, to hedge more directions in the risk space, we need more data. And such data is available \cite{AlphaFM}: the position data of the underlying tradable instruments; e.g., if our alpha is a dollar-neutral portfolio of, say, 2,000 most liquid US stocks, a time series of the positions that this alpha instructs us to take. The question is, how can we use this position data? E.g., can we improve alpha weights?

{}One idea set forth in \cite{AlphaFM} (and further discussed in \cite{Billion}) is to use the position data for the underlying tradable instruments to build risk models for alpha portfolios -- albeit without an explicit implementation. In this paper we fill this gap -- although the result is not what one would expect based on the intuition from risk modeling for stocks -- and discuss alpha risk models in detail. We do this to give an alternative -- and compelling -- way of arriving at our main result, which we first set forth without any reference to alpha portfolio risk.

{}Our idea is simple. We have a large number $N$ of alphas -- say, $N$ = 1,000,000. Let us assume that these alphas are all trading the same underlying $M$ instruments -- say, $M$ = 3,000 most liquid US stocks.\footnote{\, Each alpha may have its own universe of stocks. What is important here is that their universes have a substantial overlap. If they do not, we can always exclude the alphas with small overlap.} To determine the weights with which individual alphas contribute to the ``mega-alpha" (i.e., the alpha portfolio), we need expected returns for individual alphas. Then we can ask the following question:

{}{\em Can we forecast expected returns for the underlying tradables (stocks) using the expected returns for alphas?} Put differently, can we decode the stock market using alphas? The answer is affirmative. And this is where the position data comes in.

{}Now, if we can forecast stock expected returns using alpha expected returns, then we no longer need to combine alphas. We can trade a stock portfolio directly based on stock expected returns. And we can do all the risk management directly on this stock portfolio as opposed to doing it in two steps, by first managing the alpha portfolio risk, and then managing the risk of the stock portfolio corresponding to the ``mega-alpha". I.e., we will not need the extra step of constructing the alpha portfolio -- and intuitively it is clear that a priori this extra layer could be a source of additional noise and suboptimality. So, we should just get rid of it if we can.\footnote{\, Thereby turning an underconstrained $T \ll N$ problem into a much easier $N \gg M$ problem.}

{}In Section \ref{sec.2} we give an explicit algorithm for extracting stock expected returns from alpha expected returns. We give the source code for this algorithm in Appendix \ref{app.A}.\footnote{\, The source code given in Appendix \ref{app.A} hereof is not written to be ``fancy" or optimized for speed or in any other way. Its sole purpose is to illustrate the algorithms described in the main text in a simple-to-understand fashion. Some important legalese is relegated to Appendix \ref{app.B}.} The stock expected returns are nothing but coefficients in a weighted regression of alpha expected returns over the position data. Subtleties arise when the individual alpha portfolios are subject to linear constraints, e.g., dollar neutrality or sector/(sub-)industry neutrality, and we discuss how to deal with them. We also discuss how to choose the regression weights (other than inverse alpha variances).

{}In Section \ref{sec.3} we give an explicit algorithm for building risk models for alpha portfolios using the position data for the underlying tradables. The source code therefor is subsumed in Appendix \ref{app.A}. Following \cite{Billion}, we then show that, when $N \gg 1$, optimization using this risk model reduces to the weighted regression in Section \ref{sec.2}. A neat subtlety is that in the ``leading" order in the powers of $1/N$ the stock expected returns vanish, and it is the ``next-to-leading" order that reproduces the results of Section \ref{sec.2}. We briefly conclude in Section \ref{sec.4}.

\section{Stock Returns from Alphas}\label{sec.2}

{}For definiteness -- and the underlying instruments are not critical here -- let us focus on alphas that trade (largely) overlapping portfolios of US stocks. Let the number of all stocks traded be $M$, and let the number of alphas be $N \gg M$. Let the realized returns for alphas be $\rho_{is}$ ($i=1,\dots,N$), and the realized returns for stocks be $R_{As}$ ($A=1,\dots,M$), where the index $s=1,\dots,T$ labels trading days (for definiteness, let $s=1$ correspond to the most recent date). Then we have
\begin{equation}\label{ex-post}
 \rho_{is} = \sum_{A=1}^M P_{iAs}~R_{As}
\end{equation}
Here on each day labeled by $s$ for each alpha labeled by $i$ the quantities $P_{iAs}$ are nothing but the properly normalized stock positions in the corresponding alpha portfolio. The normalization condition is given by
\begin{equation}
 \sum_{A=1}^M |P_{iAs}| = 1
\end{equation}
The stock positions $P_{iAs}$, a.k.a. the desired holdings, are ``previsible", i.e., they are known in advance of commencing the trading that is supposed to achieve these positions. Under the hood, alphas analyze historical and real-time data up to the time these positions are calculated. Another way of phrasing this is that the positions $P_{iAs}$ are known out-of-sample. This has implications for stock expected returns.

\subsection{How to Extract Stock Expected Returns?}

{}Above we discussed the relation (\ref{ex-post}) between the realized returns for stocks $R_{As}$ and the realized returns for alphas $\rho_{is}$. These are ex-post returns. These returns are not known in advance of commencing the trading that is supposed to achieve the positions $P_{iAs}$. What is known ex-ante, i.e., what is forecast based on the historical data, are {\em expected} returns for alphas, call them $\eta_{is}$. These expected returns are also ``previsible" (or out-of-sample) by construction. E.g., a simple way to construct them is via moving averages based on prior $d$ days' worth of realized returns:\footnote{\, We emphasize that this is only an example and there are other ways of constructing $\eta_{is}$.}
\begin{equation}
 \eta_{is} = {1\over d} \sum_{s^\prime = s + 1}^{s + d} \rho_{is^\prime}
\end{equation}
In this example -- and this is a reasonable approach -- the bet is that, if an alpha on average has made money for the past $d$ days (say, $d=10$), then we expect (i.e., hope) that it should (on average) continue to make money moving forward, at least for the foreseeable future (and this expectation, as discussed above, is ephemeral). This is an example of a ``momentum" strategy (in this case for alphas, not stocks).

{}So, suppose, one way or another, we have computed our alpha expected returns $\eta_{is}$. Can we use them to reasonably define -- call them $E_{As}$ -- expected returns for stocks? Sure we can, via a linear model:
\begin{equation}\label{ex-ante}
 \eta_{is} = \epsilon_{is} + \sum_{A=1}^M P_{iAs}~E_{As}
\end{equation}
I.e., we mimic (\ref{ex-post}), except that here, for each date $s$, we have $N$ datapoints $\eta_{is}$ and many fewer $M$ unknowns $E_{As}$. The fit (\ref{ex-ante}) via a linear model has errors $\epsilon_{is}$. And a standard way to fix $E_{As}$ is by minimizing the least squares of these errors (for each value of $s$):\footnote{\, Here -- again, for each value of $s$ -- the minimization is w.r.t. the $M$-vector $E_{As}$.}
\begin{equation}\label{error}
 \sum_{i=1}^N \epsilon_{is}^2 \rightarrow \mbox{min}
\end{equation}
This is the same as running -- for each value of $s$ -- a linear cross-sectional (i.e., across the index $i$) regression of $\eta_{is}$ over the $N\times M$ loadings matrix $P_{iAs}$ (without the intercept and with unit weights -- see below). Then $\epsilon_{is}$ are the residuals of this regression, whereas $E_{As}$ are the regression coefficients given by
\begin{equation}
 E_{As} = \sum_{i=1}^N \sum_{B=1}^M Y_{ABs}~P_{iBs}~\eta_{is}
\end{equation}
where for each value of $s$ the $M\times M$ matrix $Y_{ABs}$ is the inverse of the $M\times M$ matrix
\begin{equation}
 X_{ABs} = \sum_{i=1}^N P_{iAs}~P_{iBs}
\end{equation}
So, {\em assuming} this matrix is invertible, we can calculate the expected returns for stocks $E_{As}$ using the expected returns for alphas $\eta_{is}$. The invertibility assumption may not necessarily hold and we discuss how to deal with this below in detail.

\subsubsection{Stock Portfolio}

{}However, let us for now assume the invertibility of $X_{ABs}$. Then, as advertised above, we have extracted the expected returns for stocks $E_{As}$ and we can {\em directly} construct a stock portfolio we wish to trade without any reference to a ``mega-alpha", alpha weights, alpha combos, or alpha portfolios. Thus, if we have a good risk model for our stock portfolio, i.e., a positive-definite (and thus invertible) and reasonably stable $M\times M$ risk model covariance matrix $\Phi_{AB}$ for stocks,\footnote{\, We discuss how to construct this risk model below.}, then, now that we have the stock expected returns $E_{As}$, we can construct our stock portfolio weights $w_{As}$ via, e.g., maximizing its Sharpe ratio \cite{Sharpe94}:\footnote{\, Here $\Phi_{AB}^{-1}$ is the matrix inverse to $\Phi_{AB}$. More generally, we can have a different covariance matrix $\Phi_{ABs}$ for each date $s$ in (\ref{stock.wts}) -- albeit this can increase noise. This is not critical here.\label{fn.Phi}}
\begin{equation}\label{stock.wts}
 w_{As} = \gamma~\sum_{B=1}^M \Phi_{AB}^{-1}~E_{As}
\end{equation}
where the overall normalization coefficient $\gamma$ is fixed via the normalization condition
\begin{equation}\label{stock.norm}
 \sum_{A=1}^M |w_{As}| = 1
\end{equation}
We can further add bells and whistles to our stock portfolio optimization, such as position and liquidity bounds on the stock weights $w_{As}$, dollar-neutrality and other linear constraints, as well as any other standard risk management conditions (which we will not delve into it here). The key is fixing the stock expected returns $E_{As}$.

\subsubsection{Weighted Regression}

{}Speaking of which, above we fixed them by minimizing the least squares (\ref{error}). While this is a reasonable approach, it may not be optimal. The reason why is that the realized alpha returns $\rho_{is}$ and thereby expected alpha returns $\eta_{is}$ have cross-sectionally skewed historical volatilities \cite{4K}, \cite{101}. This then invariably implies that the residuals $\epsilon_{is}$ also have cross-sectionally skewed historical volatilities. So, (\ref{error}) receives oversized contributions from volatile alphas. This can be rectified by running least-squares on appropriately normalized errors:
\begin{equation}\label{error.w}
 \sum_{i=1}^N {\widetilde \epsilon}_{is}^2 = \sum_{i=1}^N v_{is}~\epsilon_{is}^2 \rightarrow \mbox{min}
\end{equation}
Here ${\widetilde \epsilon}_{is} = \epsilon_{is} / \xi_{is}$ and $v_{is} = 1/\xi_{is}^2$. The normalizations $\xi_{is}$ should be chosen such that they take care of the aforementioned skewness. The simplest choice is to use historical alpha volatilities $\xi_{is} = \sigma_{is}$, where $\sigma_{is}$ can be computed, e.g., based on prior $d$ days' worth of historical data ($\mbox{Var}(\cdot, d)$ is a $d$-day moving {\em serial} variance):
\begin{eqnarray}\label{var.d}
 &&\sigma_{is}^2 = \mbox{Var}(\eta_{is}, d) = {1\over{d-1}}\sum_{s^\prime = s+1}^{s+d} (\eta_{is^\prime} - {\overline \eta}_{is})^2\\
 &&{\overline \eta}_{is} = {1\over d}\sum_{s^\prime = s+1}^{s+d} \eta_{is^\prime}
\end{eqnarray}
Here two questions arise naturally. First, should we compute $\sigma_{is}^2$ based on the expected returns $\eta_{is}$ or the realized returns $\rho_{is}$? Second, should we set $\xi_{is} = \sigma_{is}$ or should we compute $\xi_{is}$ some other way, e.g., as volatilities of the residuals rather than of the (expected or realized) returns for alphas? We will come back to this below. For now let us simply assume that we have some way of computing $\xi_{is}$.

{}Then (\ref{error.w}) is equivalent (for each value of $s$) to running a weighted regression (without the intercept) of $\eta_{is}$ over the loadings matrix $P_{iAs}$ with the weights $v_{is}$. The stock expected returns $E_{As}$ are the regression coefficients given by
\begin{equation}\label{v.E}
 E_{As} = \sum_{i=1}^N \sum_{B=1}^M Y_{ABs}~v_{is}~P_{iBs}~\eta_{is}
\end{equation}
where for each value of $s$ the $M\times M$ matrix $Y_{ABs}$ is the inverse of the $M\times M$ matrix
\begin{equation}\label{v.X}
 X_{ABs} = \sum_{i=1}^N v_{is}~P_{iAs}~P_{iBs}
\end{equation}
As above, invertibility of $X_{ABs}$ is not given. But first we discuss regression weights.

\subsubsection{Regression Weights}

{}So, what should the regression weights be? Since these weights are meant to normalize the regression residuals in (\ref{error.w}), it is natural to set $\xi_{is}^2$ to moving serial variances of the time series of the regression residuals $\epsilon_{is}$. However, let us first discuss why using historical alpha volatilities, which would be a simple choice, may be suboptimal.

{}Let us first look at historical volatilities of the realized alpha returns (\ref{ex-post}). When computing serial variances, we would be muddling up {\em non-factorized} contributions from the realized stock returns $R_{As}$ and the fact that the desired holdings $P_{iAs}$ also change from one date $s$ to another. However, this position data is ``previsible", it is fixed by how the alphas are constructed, and should not contribute into the measure of uncertainty (i.e., volatility) we use to normalize the regression residuals. On the other hand, the realized alpha returns know nothing about the uncertainty stemming from forecasting the alphas, which is encoded in the time series of the expected alpha return (\ref{ex-ante}). So, using $\eta_{is}$ is more reasonable, but we still must not muddle up the weights based on $\eta_{is}$ with the $s$-dependence of the desired holdings $P_{iAs}$. And this is achieved precisely by regressing $\eta_{is}$ over $P_{iAs}$ (for each date $s$), which produces the regression residuals $\epsilon_{is}$, and setting
\begin{equation}
 \xi^2_{is} = \mbox{Var}(\epsilon_{is}, d)
\end{equation}
$\mbox{Var}(\cdot, d)$ is defined in (\ref{var.d}). So, the regression with unit weights fixes $v_{is} = 1/\xi_{is}^2$.

\subsection{Linear Constraints}

{}There may be additional linear constraints on stock positions stemming from all alphas being subject to the same risk management restrictions. E.g., suppose all alphas are dollar-neutral. Then we have:
\begin{equation}\label{dollar.neutral}
 \sum_{A=1}^M P_{iAs} \equiv 0
\end{equation}
More generally, we can have $p$ linear constraints (typically, $p \ll M$, so we assume that this holds)
\begin{equation}\label{lin.con}
 \sum_{A=1}^M P_{iAs}~Q_{A\alpha} \equiv 0,~~~\alpha = 1,\dots,p
\end{equation}
Without loss of generality we can assume that: i) the $p$ columns of the matrix $Q_{A\alpha}$ are linearly independent;\footnote{\, The constraints (\ref{lin.con}) are invariant under $SO(p)$ rotations $Q\rightarrow Q~U$, where $U_{\alpha\beta}$ is an orthogonal matrix: $U U^T = U^T U = 1$.} and ii) the constraints (\ref{lin.con}) do not imply that $P_{iAs} \equiv 0$ for any given value of $A$ (or else none of the alphas trade the stock labeled by $A$ and we can simply drop it out of the universe). Furthermore, to keep it simple, we will assume that if there is a subset of $N_1$ alphas ($N_1 < N$) for which -- but not for all alphas -- we have additional linear constraints other than (\ref{lin.con}), then $N_1 \ll N$.

{}If we have $p > 0$ linear constraints, our alphas do not depend on $p$ linear combinations of the stock returns. We can decompose
\begin{eqnarray}
 &&R_{As} = R^\prime_{As} + \sum_{\alpha=1}^p Q_{A\alpha}~R_{\alpha s}\\
 &&\sum_{A=1}^M Q_{A\alpha}~R^\prime_{As} \equiv 0\label{Rprime}
\end{eqnarray}
where
\begin{eqnarray}
 &&R_{\alpha s} = \sum_{A = 1}^M {\widetilde Q}_{\alpha A}~R_{As}
\end{eqnarray}
and (in matrix notations) ${\widetilde Q} = (Q^T Q)^{-1}Q^T$. Note that the matrix $Q^TQ$ is nonsingular as the columns of $Q_{A\alpha}$ are linearly independent. So, the alpha returns
\begin{equation}
 \rho_{is} = \sum_{A=1}^M P_{iAs}~R^\prime_{As}
\end{equation}
know nothing about the returns $R_{\alpha s}$ for the {\em risk factors} defined by the {\em factor loadings} matrix $Q_{A\alpha}$. E.g., in the case of a single dollar-neutrality constraint (\ref{dollar.neutral}) we have $Q_{A\alpha} \equiv 1$ ($\alpha=1$), and the corresponding return $R_{\alpha s} = {1\over M}\sum_{A=1}^M R_{As}$ is nothing but the average stock portfolio return, which for large $M$ can be thought of as an equally-weighted broad-market return. Dollar neutrality hedges against market risk.

{}Based on the foregoing, intuitively it is clear that in the presence of linear constraints (\ref{lin.con}) we cannot forecast all stock expected returns using alpha expected returns, but only $M - p$ linear combinations thereof as the alpha portfolios are neutral under the remaining $p$ linear combinations. So, we need to reduce the number of stock returns from $M$ to $M-p$. There are various ways of achieving this goal.

\subsection{Elimination Method}

{}Thus, we can build a portfolio of $M-p$ stocks as follows. At each step $r \leq M$ we have three sets $\Pi_r$, ${\widetilde\Pi}_r$ and $\Upsilon_r$. At step $r=1$ we start with $\Pi_1 = \{A|A=1\}$, ${\widetilde \Pi}_1 =\{A|A=2,3,\dots,M\}$, and $\Upsilon_1 = \{\alpha|\alpha=1,\dots,p\}$. At each step $r$, where $r < M$, we define $\Pi^\prime_r = \Pi_r \cup \{B\}$, where $B = \mbox{min}({\widetilde\Pi}_r)$. If there is no $\beta \in \Upsilon_r$ such that
\begin{equation}
 \sum_{A\in \Pi^\prime_r} P_{iAs}~Q_{A\beta} \equiv 0
\end{equation}
then we define $\Pi_{r+1} = \Pi^\prime_r$ (i.e., we add $B$ to $\Pi_r$), ${\widetilde\Pi}_{r+1} = {\widetilde \Pi}_r \backslash \{B\}$ (i.e., we delete $B$ from ${\widetilde \Pi}_r$) and $\Upsilon_{r+1} = \Upsilon_r$. If such $\beta$ does exist, then we define $\Pi_{r+1} = \Pi_r$, ${\widetilde\Pi}_{r+1} = {\widetilde \Pi}_r \backslash \{B\}$ and $\Upsilon_{r+1} = \Upsilon_r \backslash \{\beta\}$ (i.e., we delete $\beta$ from $\Upsilon_r$). At step $r=M$ we end up with the subset $\Pi_M \subset \{1,\dots,M\}$ such that there are no linear constraints on $P_{iAs}$ for $A\in \Pi_M$. The subset $J = \Pi_M$ has $|J| = M - p$ elements. Its complement ${\widetilde J} = \{1,\dots,M\}\backslash J$ has $|{\widetilde J}| = p$ elements. Moving forward we will use the lower case characters $a,b, \dots$ (from the beginning of the Latin alphabet) to label the stocks corresponding to the subset $J$, and the characters $\mu,\nu,\dots$ (from the middle of the Greek alphabet)\footnote{\, Not to be confused with the characters $\alpha,\beta,\dots$ from the beginning of the Greek alphabet we use in $Q_{A\alpha}$.} to label the stocks corresponding to the subset ${\widetilde J}$. Then we can rewrite (\ref{Rprime}) via
\begin{equation}
 \sum_{\mu\in {\widetilde J}} q_{\mu\alpha}~R^\prime_{\mu s} = - \sum_{a\in J} K_{a\alpha}~R^\prime_{as}
\end{equation}
where (in matrix notations) $q_{\mu\alpha} = Q_{\mu\alpha}$ and $K_{a\alpha} = Q_{a\alpha}$. Therefore, we have
\begin{eqnarray}\label{R.mu}
 &&R^\prime_{\mu s} = - \sum_{a\in J} \chi_{\mu a}~R^\prime_{as}\\
 &&\rho_{is} = \sum_{a\in J} S_{ias}~R^\prime_{as}\label{rho.red}
\end{eqnarray}
where (in matrix notations) $\chi = (qq^T)^{-1}qK^T$ and
\begin{equation}
 S_{ias} = P_{ias} - \sum_{\mu\in {\widetilde J}} P_{i\mu s}~\chi_{\mu a}
\end{equation}
So, (\ref{rho.red}) now looks like there are no linear constraints, except that the stock universe is smaller ($M-p$ stocks instead of $M$), the stock ``positions" are now given by $S_{ias}$ (instead of $P_{iAs}$), and the $M-p$ returns are $R^\prime_{as}$. The remaining $p$ returns $R^\prime_{\mu s}$ are fixed via (\ref{R.mu}), and the $M$ returns $R^\prime_{As}$ are not the same as the original returns $R_{As}$ but are projected onto an $(M-p)$-dimensional hyperplane via (\ref{Rprime}). The directions $R_{\alpha s}$ perpendicular to this hyperplane are unattainable, but we can solve for $R^\prime_{As}$.

\subsection{Principal Components}\label{sub.pc}

{}There is a formally simpler method for dealing with linear constraints. The practical issue is that (for each value of $s$) the $M\times M$ matrix (\ref{v.X}) is singular in the presence of linear constraints (\ref{lin.con}). This can be dealt with as follows. Let us decompose the matrix $X_{ABs}$ using its principal components $V^{(C)}_{As}$:
\begin{equation}
 X_{ABs} = \sum_{C = 1}^M \lambda^{(C)}_s~V^{(C)}_{As}~V^{(C)}_{Bs}
\end{equation}
Let us label the positive eigenvalues via $\lambda^{(a)}_s$, $a\in J$, $|J|=M-p$, and the null eigenvalues via $\lambda^{(\mu)}_s$, $\mu\in {\widetilde J}$, $|{\widetilde J}|=p$. We can regularize the matrix $X_{ABs}$ via
\begin{equation}
 X_{ABs} = \sum_{a\in J} \lambda^{(a)}_s~V^{(a)}_{As}~V^{(a)}_{Bs} + \sum_{\mu\in {\widetilde J}} \lambda_0~V^{(\mu)}_{As}~V^{(\mu)}_{Bs}
\end{equation}
where at the end of the day we will take $\lambda_0\rightarrow 0+$. Note that, by definition, we have
\begin{eqnarray}
 &&\sum_{A=1}^M V^{(a)}_{As}~Q_{A\alpha} \equiv 0\\
 &&\sum_{A=1}^M V^{(\mu)}_{As}~P_{iAs} \equiv 0
\end{eqnarray}
The inverse of $X_{ABs}$ is given by
\begin{equation}
 Y_{ABs} = \sum_{a\in J} [\lambda^{(a)}_s]^{-1}~V^{(a)}_{As}~V^{(a)}_{Bs} + \sum_{\mu\in {\widetilde J}} \lambda_0^{-1}~V^{(\mu)}_{As}~V^{(\mu)}_{Bs}
\end{equation}
Plugging this into (\ref{v.E}), we get
\begin{equation}\label{stock.exp.ret.con}
 E_{As} = \sum_{i=1}^N \sum_{B=1}^M \sum_{a\in J} [\lambda^{(a)}_s]^{-1}~V^{(a)}_{As}~V^{(a)}_{Bs}~v_{is}~P_{iBs}~\eta_{is}
\end{equation}
This expression is independent of the regulator $\lambda_0$, which we can safely take to 0. Note that
\begin{equation}
 \sum_{A=1}^M E_{As}~Q_{A\alpha} \equiv 0
\end{equation}
Once again, only the directions orthogonal to the linear constraints are attainable.

\section{Alpha Risk Models}\label{sec.3}

{}The formulas (\ref{stock.exp.ret.con}), (\ref{stock.wts}) and (\ref{stock.norm}) express the stock portfolio weights via alpha expected returns, without any reference to alpha portfolio weights. We will now derive this result in a seemingly unrelated way, by constructing risk models for alpha portfolios.

{}So, as it is normally done in practice, let us combine our $N$ alphas with some weights $w_{is}$, which we need to fix somehow. Let us look at the risk of the underlying stock portfolio corresponding to these alpha weights. The corresponding stock weights are given by
\begin{equation}\label{stock.alpha}
 w_{As} = \sum_{i = 1}^N P_{iAs}~w_{is} 
\end{equation} 
On a given date $s$, the (expected) stock portfolio variance is given by
\begin{eqnarray}
 &&\sum_{A,B = 1}^M \Phi_{AB}~w_{As}~w_{Bs} = \sum_{i,j = 1}^N w_{is}~w_{js}~F_{ijs}\\
 &&F_{ijs} = \sum_{A,B=1}^M P_{iAs}~\Phi_{AB}~P_{jBs} \label{F}
\end{eqnarray}
where, as above, $\Phi_{AB}$ is a positive-definite (and reasonably stable) $M \times M$ risk model covariance matrix for stocks.\footnote{\, The comment in footnote \ref{fn.Phi} applies to (\ref{F}) as well.} Even though $\Phi_{AB}$ is invertible, the $N\times N$ matrix $F_{ijs}$ (for each value of $s$) is singular -- there are many more alphas than stocks. In fact, the l.h.s. is nothing but an incomplete {\em factor model} with $M$ factors, the factor loadings matrix $P_{iAs}$, and the factor covariance matrix $\Phi_{AB}$. What is missing is the specific (a.k.a. idiosyncratic) risk on the diagonal. Once we add the specific risk, call it $\zeta_{is}^2$, we have an $M$-factor model with a positive-definite model covariance matrix for our $N$ alphas:\footnote{\, We can think about the specific risk in the matrix $\Gamma_{ijs}$ in (\ref{Gamma}) as modeling the uncertainty in the alpha expected returns other than that due to stock volatility. The latter is modeled by the factor risk $F_{ijs}$. We will come back to addressing whether the former should be diagonal below.}
\begin{equation}\label{Gamma}
 \Gamma_{ijs} = \zeta_{is}^2~\delta_{ij} + F_{ijs}
\end{equation} 
We will discuss what $\zeta_{is}$ should be in a moment. For now, let us assume we know how to compute them and figure out what the weights $w_{is}$ are based on the alpha expected returns $\eta_{is}$ and the alpha portfolio risk modeled by $\Gamma_{ijs}$. As for stocks, let us fix $w_{is}$ by maximizing the Sharpe ratio \cite{Sharpe94} of the alpha portfolio:
\begin{equation}\label{alpha.w}
 w_{is} = \kappa \sum_{j=1}^N \Gamma_{ijs}^{-1}~\eta_{js}
\end{equation}
Here, for each date $s$, $\Gamma_{ijs}^{-1}$ is the $N\times N$ matrix inverse to $\Gamma_{ijs}$, and the overall normalization coefficient $\kappa$ is fixed via the normalization condition
\begin{equation}
 \sum_{i=1}^N |w_{is}| = 1
\end{equation}
We will see momentarily that the alpha weights $w_{is}$ reduce to a familiar form.

\subsection{Large $N$ Limit}

{}For our purposes here it is convenient to rewrite $\Gamma_{ijs}$ via $\Gamma_{ijs} = \zeta_{is}~\zeta_{js}~\gamma_{ijs}$, where
\begin{equation}
 \gamma_{ijs} = \delta_{ij} + \sum_{A=1}^M \beta_{iAs}~\beta_{jAs}
\end{equation}
and $\beta_{iAs} = {\widetilde \beta}_{iAs} / \zeta_{is}$. Here ${\widetilde\beta}_{iAs} = \sum_{B=1}^M P_{iBs}~\phi_{BA}$, and $\phi$ is the Cholesky decomposition of $\Phi$, so (in matrix notations) $\phi~\phi^T = \Phi$. We then have
\begin{equation}\label{w.i}
 w_{is} = {\kappa\over\zeta_{is}}\sum_{j = 1}^N \gamma_{ij}^{-1}~{\eta_{js}\over\zeta_{js}} = {\kappa\over\zeta_{is}}\left[{\eta_{is}\over\zeta_{is}} - \sum_{j = 1}^N \sum_{A,B = 1}^M \beta_{iAs}~Q^{-1}_{ABs}~\beta_{jBs}~{\eta_{js}\over\zeta_{js}}\right]
\end{equation}
where (for each date $s$) $Q^{-1}_{ABs}$ is the $M\times M$ matrix inverse to $Q_{ABs} = \delta_{AB} + q_{ABs}$, and $q_{ABs} = \sum_{i=1}^N \beta_{iAs}~\beta_{iBs}$. The diagonal elements of this matrix are $Q_{AAs} = 1 + \sum_{i=1}^N \beta_{iAs}^2$. In a moment we will argue that all $q_{AAs} = \sum_{i=1}^N \beta_{iAs}^2 \gg 1$. Then we can expand $Q^{-1}_{ABs}$ as follows:
\begin{equation}\label{Q.inv}
 Q^{-1}_{ABs} = q^{-1}_{ABs} - \sum_{C=1}^M q^{-1}_{ACs}~q^{-1}_{CBs} + {\cal O}(q^{-3})
\end{equation}
Here (for each date $s$) $q^{-1}_{ABs}$ is the $M\times M$ matrix inverse to $q_{ABs}$. The first term in (\ref{Q.inv}) gives the leading contribution of order $1/N$ into $Q^{-1}_{ABs}$, the second term is the next-to-leading contribution of order $1/N^2$, and the remaining terms in the expansion, which we schematically denoted as ${\cal O}(q^{-3})$, are of order $1/N^3$, $1/N^4$, etc. As we will see in a moment, we need to keep not just the leading term $q^{-1}_{ABs}$, but also the next-to-leading (second) term in (\ref{Q.inv}), and we can safely discard the rest.\footnote{\, One caveat here is that $q_{ABs}$ is (for now) assumed to be invertible. We will relax this below.}

{}So, why are all $q_{AA}\gg 1$? This is the case when \cite{Billion}: i) $N$ is large, and ii) there is no ``clustering" in the vectors $\beta_{iA}$. That is, we do not have vanishing or small values of $\beta_{iA}^2$ for most values of the index $i$ with only a small subset thereof having $\beta_{iA}^2\gsim 1$. Without such ``clustering", which is not observed in practice, to have $q_{AA}\lsim 1$, we would have to have $\beta^2_{iA}\ll 1$, i.e., $\gamma_{ij}$ and, therefore, $\Gamma_{ij}$ would be almost diagonal. Such a risk model would not describe realistic alphas.\footnote{\, In practice, alphas are not too highly correlated -- this is because one does not wish to trade highly correlated alphas. However, empirically the average correlation between alphas is not tiny either (definitely, not of order $1/N$) \cite{101}.}

\subsection{Stock Portfolio}

{}Let us now plug (\ref{w.i}) and (\ref{Q.inv}) into (\ref{stock.alpha}). Straightforward algebra yields:
\begin{equation}\label{w.fin}
 w_{As} = \kappa\sum_{i=1}^N\sum_{B,C = 1}^M \Phi_{AB}^{-1}~{\widetilde Y}_{BCs}~{\widetilde v}_{is}~P_{iCs}~\eta_{is} + \dots
\end{equation}
where the ellipses stand for subleading terms corresponding to the ${\cal O}(q^{-3})$ contributions in (\ref{Q.inv}), ${\widetilde v}_{is} = 1/\zeta_{is}^2$, and ${\widetilde Y}_{ABs}$ is the inverse of the matrix
\begin{equation}\label{X.tilde}
 {\widetilde X}_{ABs} = \sum_{i=1}^N {\widetilde v}_{is}~P_{iAs}~P_{jAs}
\end{equation}
I.e., here we have the exact same result as that given by (\ref{v.E}), (\ref{stock.wts}) and (\ref{stock.norm}) provided that we identify $\zeta_{is}$ with $\xi_{is}$. More precisely, this identification is up to an overall ($s$-dependent) constant, which does not affect the final result. This overall normalization constant plays the role of the relative normalization between the specific risk and factor risk in (\ref{Gamma}) and is nontrivial to fix as it depends on the normalization used in constructing the stock risk model covariance matrix $\Phi_{AB}$. However, the beauty of the large $N$ limit is that we do not need to know this normalization constant as it only affects the subleading terms suppressed by powers of $1/N$ (i.e., the terms corresponding to the ellipses in (\ref{w.fin})). Also, note that the leading contribution in (\ref{w.fin}) comes from the next-to-leading (second) term in (\ref{Q.inv}). There is a simple reason for this. Suppose we kept only the leading (first) term in (\ref{Q.inv}). We would then get $w_{As}\equiv 0$. Indeed, keeping the leading term in (\ref{Q.inv}) reduces (\ref{w.i}) to a weighted cross-sectional regression \cite{Billion} of $\eta_{is}$ over $P_{iAs}$ (with the weights $1/\zeta_{is}^2$). Then $w_{is}$ are automatically orthogonal to $P_{iAs}$, so $w_{As}$ (given by (\ref{stock.alpha})) automatically vanish, even though $w_{is}$ do not! This is because the alpha portfolio is perfectly hedged against the risk factors, which are nothing but the stock returns. So, the combined alpha portfolio has exactly zero stock holdings. To obtain a nontrivial stock portfolio, we must relax this perfect hedge, i.e., move away from the exact regression to optimization. Keeping the next-to-leading term in (\ref{Q.inv}) accomplishes exactly that. And, happily, precisely because $N$ is large, all the other terms are suppressed and we do not even need to fix the relative normalization between the specific and factor risks, which only affects these suppressed terms.

{}One loose end we need to tie up is that above we assumed that $q_{AB}$ is invertible. This is the case if we do not have linear constraints. In the presence of linear constraints (\ref{lin.con}) $q_{AB}$ is singular. It can be regularized as in Subsection \ref{sub.pc} and then the rest of the argument goes through. The net result is that
\begin{equation}\label{stock.wts.x}
 w_{As} = \kappa\sum_{i=1}^N \sum_{B,C=1}^M \sum_{a\in J} \Phi^{-1}_{AB}~[\lambda^{(a)}_s]^{-1}~V^{(a)}_{Bs}~V^{(a)}_{Cs}~v_{is}~P_{iCs}~\eta_{is} +\dots
\end{equation}
where the notations are the same as in Subsection \ref{sub.pc} (ellipses = subleading terms).

\subsection{A Tweak}

{}In the alpha risk model (\ref{Gamma}) we treat the position data $P_{iAs}$ as the factor loadings matrix (for each date $s$) and thereby stock returns are identified with the risk factor returns. The remainder of the risk is assumed to be diagonal specific risk, and $\zeta_{is}^2$ are identified -- {\em up to an overall ($s$-dependent) normalization factor}, which we do not need to compute for the reasons discussed above\footnote{So, without affecting the final result, here we will set it to 1 for notational simplicity. This is equivalent to assuming that the stock risk model covariance matrix $\Phi_{AB}$ is properly normalized, albeit, once again, this overall normalization is immaterial at the end of the day.} -- with the $d$-day moving variances $\xi^2_{is}$ of the regression residuals (see above). This is equivalent (for each date $s$) to modeling the ($d$-day moving) sample covariance matrix
\begin{eqnarray}
 &&\Xi_{ijs} = \mbox{Cov}(\epsilon_{is}, \epsilon_{js}, d) = {1\over {d-1}}\sum_{s^\prime=s+1}^{s+d} \left(\epsilon_{is^\prime} - {\overline\epsilon}_{is}\right)
 \left(\epsilon_{js^\prime} - {\overline\epsilon}_{js}\right)\\
 &&{\overline\epsilon}_{is} = {1\over d}\sum_{s^\prime = s+1}^{s+d} \epsilon_{is^\prime}
\end{eqnarray} 
of the regression residuals $\epsilon_{is}$ via a diagonal matrix $\mbox{diag}(\Xi_{ijs}) = \Xi_{iis}~\delta_{ij} = \xi_{is}^2~\delta_{ij}$. There are other possibilities here. We can approximate $\Xi_{ijs}$ via a non-diagonal matrix ${\widetilde \Xi}_{ijs}$ so long as it is positive-definite. Generally, the rank of the $N\times N$ matrix $\Xi_{ijs}$ equals $d-1 \ll N$. Let its principal components with positive eigenvalues $\theta^{(r)}_s$ be $U^{(r)}_{is}$, $r=1,\dots,d-1$. Let us order the eigenvalues in the descending order: $\theta^{(1)}_s > \theta^{(2)}_s > \dots > \theta^{(d-1)}_s$. Then we can construct ${\widetilde \Xi}_{ijs}$ as a $K_s$-factor statistical risk model (see, e.g., \cite{StatRM}) using the first $K_s < d-1$ principal components:
\begin{eqnarray}\label{Xi.K}
 &&{\widetilde \Xi}_{ijs} = {\widetilde\xi}_{is}^2~\delta_{ij} + \sum_{r = 1}^{K_s} \theta^{(r)}_s~U^{(r)}_{is}~U^{(r)}_{js}\\
 &&{\widetilde\xi}_{is}^2 = \xi_{is}^2 - \sum_{r = 1}^{K_s} \theta^{(r)}_s~\left[U^{(r)}_{is}\right]^2
\end{eqnarray}
Alternatively, we can model the sample correlation matrix $\Psi_{ijs} = \Xi_{ijs}/\xi_{is}~\xi_{js}$ ($\Psi_{iis}\equiv 1$) via a $K_s$-factor statistical risk model:
\begin{eqnarray}
 &&{\widetilde \Xi}_{ijs} = \xi_{is}~\xi_{js}~{\widetilde\Psi}_{ijs}\\
 &&{\widetilde \Psi}_{ijs} = {\widetilde\psi}_{is}^2~\delta_{ij} + \sum_{r = 1}^{K_s} \phi^{(r)}_s~S^{(r)}_{is}~S^{(r)}_{js}\\
 &&{\widetilde\psi}_{is}^2 = 1 - \sum_{r = 1}^{K_s} \phi^{(r)}_s~\left[S^{(r)}_{is}\right]^2
\end{eqnarray}
Here $\phi^{(r)}_s$ ($r=1,\dots,d-1$) are the positive eigenvalues of the sample correlation matrix $\Psi_{ijs}$ ($\phi^{(1)}_s > \phi^{(2)}_s > \dots > \phi^{(d-1)}_s$) and $S^{(r)}_{is}$ are the corresponding eigenvectors.

{}Following \cite{StatRM}, the number of factors $K_s$ can be fixed using the effective rank (or eRank) \cite{RV}: $K_s = \mbox{floor}(\mbox{eRank}({\cal A}_{ijs}))$ or $K_s = \mbox{round}(\mbox{eRank}({\cal A}_{ijs}))$, where ${\cal A}$ stands for the sample covariance matrix $\Xi$ or the sample correlation matrix $\Psi$. Here we can simplify things a bit and have uniform $K$ for all values of $s$, e.g., $K = \mbox{min}(K_s)$. In the discussion below, for the sake of definiteness, let us assume uniform $K_s\equiv K$ in the factor models (\ref{Xi.K}) based on the principal components of the sample covariance matrix (this is not critical).

{}So, now our alpha risk model covariance matrix instead of (\ref{Gamma}) is given by
\begin{equation}\label{Gamma1}
 {\widetilde \Gamma}_{ijs} = {\widetilde\Xi}_{ijs} + F_{ijs} = {\widetilde\xi}_{is}^2~\delta_{ij} + \sum_{{\widetilde A},{\widetilde B}\in H} {\widetilde P}_{i{\widetilde A}s} ~{\widetilde\Phi}_{{\widetilde A}{\widetilde B}}~{\widetilde P}_{j{\widetilde B}s}
\end{equation}
where the set $H = \{A\}\cup\{r\}$ (so $|H| = M + K$) is the union of the values of the indices $A,B,\dots = 1,\dots,M$ and the eigenvalue label $r = 1,\dots,K$, and ${\widetilde P}_{iAs} = P_{iAs}$, ${\widetilde P}_{irs} = [\theta^{(r)}_s]^{1/2}~U^{(r)}_{is}$. Also, ${\widetilde \Phi}_{AB} = \Phi_{AB}$, ${\widetilde \Phi}_{rr^\prime} = \delta_{rr^\prime}$, and other components vanish. Now, the regression residuals $\epsilon_{is}$ by definition are orthogonal to $P_{iAs}$, so we have $\sum_{i=1}^N\epsilon_{is}~P_{iAs} \equiv 0$. This implies that $\sum_{i=1}^N U^{(r)}_{is}~P_{iAs} \equiv 0$. Straightforward algebra then yields a simple result: the stock portfolio weights are still given by (\ref{stock.wts.x}) with the only difference that now $v_{is} = 1/{\widetilde\xi}_{is}^2$ (as opposed to $v_{is} = 1/\xi_{is}^2$). I.e., the effect of modeling the matrix $\Xi_{ijs}$ via its principal components simply reduces to modifying the regression weights. So, the choice of $v_{is}$ encodes the difference between models!\footnote{\, Here one may wonder how to control the stock portfolio turnover as high turnover alphas may increase it. A simple method is to suppress the weights $v_{is}$ for high turnover alphas.\label{fn.tvr}}

\section{Concluding Remarks}\label{sec.4}

{}As mentioned above, there is an overall normalization coefficient between the specific risk and factor risk terms in (\ref{Gamma}) (and thereby also in (\ref{Gamma1})). In fact, this overall normalization generally is $s$-dependent. It is related to the fact that the normalization of the stock risk model covariance matrix $\Phi_{AB}$ a priori is not the same as that of the resultant expected stock returns. However, as discussed above, for alpha portfolio optimization purposes we do not need to determine this overall normalization. This is because we are not actually inverting $\Gamma_{ijs}$; instead, we are running a regression over the position data $P_{iAs}$ with the weights $v_{is}$ determined via the specific risks $\xi_{is}$. 

{}This is different from the risk model building for stocks,\footnote{\, For a general discussion and references, see, e.g., \cite{GK}. For a detailed construction (including important details overlooked in prior literature and commercial offerings) and explicit implementation complete with source code, see \cite{HetPlus}.} where one constructs a full multifactor risk model matrix $\Phi_{AB}$ with no undetermined normalization factors, etc. So, why is there such a glaring difference between stock and alpha risk models? The answer is simple. In stock risk models the industry-based risk factors play a dominant role, notably, by their ubiquity (as opposed to style factors or principal components -- see \cite{HetPlus}). For the industry-based risk factors the corresponding factor loadings matrix $\Omega_{AI}$ (where $I$ labels industries) does, in fact, have a ``clustering" structure; e.g., for binary industry classifications $\Omega_{AI} = 1$ if the stock labeled by $A$ belongs to the industry labeled by $I$, otherwise $\Omega_{AI} = 0$. Therefore, Sharpe ratio maximization based on such $\Phi_{AB}$ does {\em not} reduce to a regression.\footnote{\, In contrast, Shape ratio maximization using stock risk models based solely on style factors and/or principal components always reduces to a regression (assuming $M\gg 1$). This also applies to the so-called ``shrinkage" models \cite{LW}, which are special cases of principal components based models \cite{StatRM}.} In some sense, alpha risk models are ``simpler" than stock risk models.

{}Speaking of stock risk models, what should we use for $\Phi_{AB}$? Off-the-shelf commercial risk models might appear as a natural choice. However, for the reasons discussed in detail in \cite{KL}, for short-horizon trading applications custom-built risk models are preferable, especially when open source code for heterotic risk models \cite{Het} and heterotic CAPM \cite{HetPlus} is freely available. This paper for alphas is in some sense analogous to \cite{HetPlus} for stocks: it provides source code for extracting stock returns from alphas directly, bypassing ``alpha combos" altogether. The implication goes beyond by-now-standard quant trading based on ephemeral alphas. These expected returns can be used for other purposes. For instance, knowing short-horizon stock expected returns can be useful to institutional portfolio managers in gauging when to execute their (typically, large) orders and thereby potentially mitigate at least some market impact effects. In this regard it is important to keep in mind that constrained alphas yield accordingly constrained stock expected returns. E.g., if all alphas are dollar-neutral, the resulting stock expected returns are demeaned: we cannot forecast the overall movement of the broad market using dollar-neutral portfolios.\footnote{\, Notwithstanding that dollar neutrality does not necessarily equal market neutrality.} Thus, developing a large number of unconstrained alphas is warranted.

{}For quant trading, apart from reducing noise and suboptimality introduced by alpha combos, one immediate benefit of our new method is that hedge funds no longer need to pay (about 3\% of the P\&L) for alpha combos (on top of about 10\% for alphas). At the end, the difference between different models (or alpha combos) reduces to picking the regression weights $v_{is}$ (which may include, e.g., turnover suppression, etc. -- see fn. \ref{fn.tvr}). Why would anyone want to pay for that? Albeit, paraphrasing, {\em avaritia caecus est}... In any event, let us finally note that we present no backtests in this paper as the position and other data are highly proprietary.

\section*{Acknowledgements}

{}ZK would like to thank Daniele Bernardi for an invitation to give a keynote talk ``How to Combine a Billion Alphas" at Quant2017, March 2-3, Venice, Italy: walking around in beautiful Venice served as a muse and an inspiration for this paper.

\appendix

\section{R Code for Stock Expected Returns}\label{app.A}

{}In this appendix we give the R source code\footnote{\, The R Project for Statistical Computing, www.r-project.org.} for extracting stock expected returns from alpha expected returns. The code below is essentially self-explanatory and straightforward as it simply follows the algorithms and formulas in Sections \ref{sec.2} and \ref{sec.3}. The entry function is {\tt{\small stk.exp.ret(ret, hld, k, tol = 1e-8)}}. Here {\tt{\small ret}} is an $N\times d$ matrix $\eta_{is}$ of alpha expected returns, where the most recent date corresponds to $s=1$; {\tt{\small hld}} is a 3-dimensional $N\times M\times d$ array $P_{iAs}$ of stock positions; {\tt{\small k}} is the input (see below) number of principal components $K$ used in modeling the matrix ${\widetilde \Xi}_{ijs}$; {\tt{\small tol}} is used in dealing with rounding errors (to wit, to distinguish null eigenvalues). Internally the function {\tt{\small stk.exp.ret()}} calls the function {\tt{\small calc.spec.var()}}, which internally calls the function {\tt{\small calc.erank()}}. For $K=0$ or $K\geq d-2$ the code sets $K=0$, i.e., no principal components are used. For $K < 0$ the code sets $K$ using eRank (truncated to an integer -- it is straightforward to change it to rounding). For $0 < K < d-2$ the code uses this input value for the number of principal components. Internally the function {\tt{\small calc.spec.var()}} also calls the function {\tt{\small qrm.calc.eigen.eff()}} given in Appendix C of \cite{StatRM}, which provides a much more efficient method ({\em not} based on power iterations) for computing eigenpairs than the internal R function (which is based on power iterations \cite{PowerIt}).\footnote{\, For large $N$ using the internal R function {\tt{\small eigen()}} here would be computationally prohibitive.} The function {\tt{\small stk.exp.ret()}} returns the vector $E_i = E_{i,s=1}$ of the stock expected returns for the most recent date $s=1$. The code only uses data with $s > 1$ to compute the regression weights out-of-sample.\\
\\
{\tt{\small
\noindent calc.erank <- function (x, excl.first)\\
\{\\
\indent take <- x > 0\\
\indent x <- x[take]\\
\indent if(excl.first)\\
\indent \indent x <- x[-1]\\
\indent p <- x / sum(x)\\
\indent h <- - sum(p * log(p))\\
\indent er <- exp(h)\\
\indent if(excl.first)\\
\indent \indent er <- er + 1\\
\indent return(er)\\
\}\\
\\
calc.spec.var <- function (x, k, do.trunc = T)\\
\{\\
\indent m <- ncol(x) - 1\\
\indent if(k == 0 | k >= m)\\
\indent \indent return(apply(x, 1, var))\\
\\
\indent x.e <- qrm.calc.eigen.eff(x)\\
\indent x.val <- x.e\$values\\
\indent x.vec <- x.e\$vectors\\
\indent if(k < 0)\\
\indent \{\\
\indent \indent k <- calc.erank(x.val, excl.first = F)\\
\indent \indent if(do.trunc)\\
\indent \indent \indent k <- trunc(k)\\
\indent \indent else\\
\indent \indent \indent k <- round(k)\\
\indent \}\\
\\
\indent if(length(take <- (k+1):m) == 1)\\
\indent \indent return(x.vec[, take]\^{}2 * x.val[take])\\
\\
\indent return(colSums(t(x.vec[, take])\^{}2 * x.val[take]))\\
\}\\
\\
stk.exp.ret <- function (ret, hld, k, tol = 1e-8)\\
\{\\
\indent n <- nrow(ret)\\
\indent d <- ncol(ret)\\
\indent res <- matrix(NA, n, d - 1)\\
\indent for(s in 2:d)\\
\indent \indent res[, s - 1] <- residuals(lm(ret[, s] $\sim$ -1 + hld[, , s]))\\
\\
\indent v <- 1 / calc.spec.var(res, k)\\
\indent x <- t(hld[, , 1]) \%*\% (v * hld[, , 1])\\
\indent x.e <- eigen(x)\\
\indent x.val <- x.e\$values\\
\indent x.vec <- x.e\$vectors\\
\\
\indent take <- x.val > tol * x.val[1]\\
\indent x.val <- x.val[take]\\
\indent x.vec <- x.vec[, take]\\
\indent stk <- colSums(hld[, , 1] * v * ret[, 1])\\
\indent stk <- colSums(x.vec * stk) / x.val\\
\indent stk <- colSums(t(x.vec) * stk)\\
\indent return(stk)\\
\}
}}

\section{DISCLAIMERS}\label{app.B}

{}Wherever the context so requires, the masculine gender includes the feminine and/or neuter, and the singular form includes the plural and {\em vice versa}. The author of this paper (``Author") and his affiliates including without limitation Quantigic$^\circledR$ Solutions LLC (``Author's Affiliates" or ``his Affiliates") make no implied or express warranties or any other representations whatsoever, including without limitation implied warranties of merchantability and fitness for a particular purpose, in connection with or with regard to the content of this paper including without limitation any code or algorithms contained herein (``Content").

{}The reader may use the Content solely at his/her/its own risk and the reader shall have no claims whatsoever against the Author or his Affiliates and the Author and his Affiliates shall have no liability whatsoever to the reader or any third party whatsoever for any loss, expense, opportunity cost, damages or any other adverse effects whatsoever relating to or arising from the use of the Content by the reader including without any limitation whatsoever: any direct, indirect, incidental, special, consequential or any other damages incurred by the reader, however caused and under any theory of liability; any loss of profit (whether incurred directly or indirectly), any loss of goodwill or reputation, any loss of data suffered, cost of procurement of substitute goods or services, or any other tangible or intangible loss; any reliance placed by the reader on the completeness, accuracy or existence of the Content or any other effect of using the Content; and any and all other adversities or negative effects the reader might encounter in using the Content irrespective of whether the Author or his Affiliates is or are or should have been aware of such adversities or negative effects.

{}The R code included in Appendix \ref{app.A} hereof is part of the copyrighted R code of Quantigic$^\circledR$ Solutions LLC and is provided herein with the express permission of Quantigic$^\circledR$ Solutions LLC. The copyright owner retains all rights, title and interest in and to its copyrighted source code included in Appendix \ref{app.A} hereof and any and all copyrights therefor.

\end{document}